\documentclass[twocolumn,aps, prapplied,Letter, floatfix]{revtex4-2}
\usepackage{graphicx}% Include figure files
\usepackage{upgreek}        
\usepackage{amssymb}               
\usepackage{amsfonts}               
\usepackage{amsmath}               
\usepackage{txfonts}  
\usepackage{textpos}      
\usepackage{array,booktabs}
\usepackage{hyperref}
\hypersetup{colorlinks=true, citecolor=blue, urlcolor=blue, linkcolor=blue} 

\begin{document}  

\begin{textblock*}{\textwidth}(0cm,0cm)
{{\it This is the accepted author version of the manuscript published in} Physical Review Applied vol. 17, 024064 (2022) [$\copyright$ 2022 American Physical Society]. The version of record is available from the publisher's website from the link: \url{https://doi.org/10.1103/PhysRevApplied.17.024064}}
\end{textblock*}  
\vspace{4cm} 
    
\title{Breaking transmission symmetry without breaking reciprocity\\ in linear all-dielectric polarization-preserving metagratings}        
\author{S. Foteinopoulou}      
\email{sfoteino@unm.edu} 
\affiliation{Department of Electrical and Computer Engineering, University of New Mexico, New Mexico 87131-0001, USA}   
\begin{abstract}                
{Transmission asymmetry in reciprocal systems offers an appealing alternative to bulkier non-reciprocal implementations for certain applications. Common reciprocal routes to transmission asymmetry of linearly polarized light involve a rotation of its polarization. Here, we explore a different route with a linear all-dielectric metagrating that preserves polarization, while lacking inversion symmetry along the surface-normal direction. Our all-angle transmission calculations reveal an abrupt transition from a symmetric to an asymmetric transmission response that traces the Bragg critical wavelength of higher-order beam emergence as a function of the incident angle. By adopting an analogy between scattering from a multi-port network and the metagrating paradigm we establish why the only necessary condition for transmission symmetry breaking in this class of systems, is the emergence of any higher-order Bragg diffracted beam. We further show how such a transmission symmetry breaking is consistent with reciprocity and also demonstrate the underpinning symmetry-breaking mechanism with a first-principle numerical experiment. Finally, we elucidate on some previous misconceptions  regarding transmission symmetry breaking related to the role of the substrate or need for change of diffraction order number at each interface.  Our proposed metagrating can exhibit a strong transmission asymmetry, with contrast that can be as high as $\sim 75\%$, thus underlining its potential as a blueprint for passive asymmetric or non-linear self-biasing non-reciprocal metasurfaces relevant to integrated and active photonics.} \\  
\end{abstract}   
\maketitle                   
\section{Introduction}  
\par
Transmission asymmetry in photonic systems,  i.e. a transmission that depends on which side of the structure
light is incident from, is a highly desirable property in modern photonics as it opens-up additional degrees of freedom for designed beam control across the EM spectrum. Transmission asymmetry of linearly-polarized light is generally regarded as synonymous to non-reciprocity and in its ideal form implies optical isolation; a property needed, for example, in lasers to protect the beam quality against back-reflections \cite{isolator, tutorial}. Such a non-reciprocal behavior is typically onset by an external static magnetic field \cite{isolator, tutorial, caloz, veronis1} which however leads to devices with a large footprint.  A weaker form of non-reciprocity \cite{caloz, shanhui, silveirinha} can be triggered by the intensity of the impinging electromagnetic (EM) wave in non-linear material systems that are asymmetric along the propagation direction \cite{mingaleev, soljacic, powell, zheludev, lifan, engheta, sounas, alu} (self-biasing non-reciprocity \cite{caloz}). The capability of a strong transmission asymmetry in these non-linear systems, without necessarily isolation \cite{caloz, silveirinha},  is certainly winning on the trade-off of more compact implementations.  
\par    
A linear passive \cite{passive} route to transmission asymmetry is most certainly attractive due to its simplicity but appears counter-intuitive. However, many authors have reported transmission symmetry breaking of linearly polarized light without violating reciprocity in structures capable of rotating its polarization \cite{carsten, halfswas, twistlight, kenanakis, povinelli, asymyogesh, dielrot, ewold}. The phenomenon stems from a planar structural asymmetry \cite{note1, nanoph_tretyak} that is different at each of the structure's faces. Specifically, in the aforementioned systems transmission is asymmetric because $S_{yx}^{+} \neq S_{yx}^{-}$, with $S_{yx}$ representing the respective scattering matrix element for light incident with $x$- linear polarization transmitted into the $y$- linear polarization channel (90 deg. polarization rotation), along the forward direction [($+$) superscript] and reverse direction [($-$) superscript] respectively. This inequality is consistent with reciprocity that requires  $S_{yx}^{+}$ to be equal only to $S_{xy}^{-}$ but not necessarily to $S_{yx}^{-}$ \cite{caloz}.
\par         
Here, we explore an alternate reciprocal route to transmission symmetry breaking of linearly-polarized light without a rotation of its polarization in a class of metagratings. We aim to unveil a core general principle that can guide the design of practically realizable structures for transmission asymmetry relevant to applications from infrared to visible frequencies while showing its consistency with reciprocity. Our study focuses on a class of linear, passive, polarization-preserving metagrating (LPPPMG) systems. It is assumed the metagrating systems of this class comprise only isotropic optical materials and may include a substrate but would not include any elements made of photonic-band-gap (PBGs) \cite{joannbook, sfanom} or hyperbolic media \cite{hyperbolic}. While transmission asymmetry can be engineered by structuring media like PBGs which possess certain directions of forbidden propagation (e.g. see Ref. \onlinecite{hamza}), these type of systems are out of the scope and discussions in this work. 
\par
Although some previous works have reported transmission asymmetry for linearly polarized light with systems falling under the LPPPMG class there is no clear consensus over the key underpinning mechanism. Specifically, an observed transmission asymmetry has been attributed to having structural elements of a different lateral periodicity \cite{cascgrat}, to a different number of diffraction orders at each side of the structure \cite{lockyear}, or to the Raleigh-Wood anomaly \cite{aydin} with dependence on the substrate's dielectric, suggesting the excitation of surface waves as the key mechanism \cite{wood}. Other works \cite{hamza2, concita} allude to the periodic pitch being the crucial controlling parameter but with no or true explanation on how the periodic pitch relates to the cause of transmission symmetry breaking. Additionally, the observed transmission-asymmetry onset for the case of normally incident light was correlated to the emergence of two first-order Bragg diffracted beams in Ref. \onlinecite{optcom} but without an explanation as to why these Bragg beams \cite{russell, zengerle} enable the transmission symmetry breaking. Hence, these previous works neither have established a general consistent transmission symmetry breaking principle for linearly polarized light applicable to the LPPPMG class nor have they provided an explanation why the observed transmission symmetry breaking does not violate reciprocity. This will be the focus of this paper. 
\par  
In particular, this paper is organized as follows. In Sec. II we perform an all-angle transmission study through an all-dielectric paradigm of the LPPPMG class and show it can exhibit a high degree of transmission asymmetry without optical isolation \cite{isolator}. We observe an abrupt phase transition from a symmetric-transmission to an asymmetric-transmission phase exactly at the Bragg critical wavelength where higher-order beams start to emerge for a certain incident angle. In Sec. III, we consider a multi-port-network framework and demonstrate that the emergence of at least one higher-order Bragg beam constitutes the necessary condition for transmission symmetry breaking in LPPPMG systems with no inversion symmetry along the propagation direction, while keeping reciprocity intact. Furthermore, in Sec. IV, we present the results of a numerical experiment on the metagrating paradigm of Sec. II with the Finite Difference Time Domain (FDTD) method \cite{taflove}, thus visualizing the mechanism of the transmission-symmetry breaking phenomenon discussed in Sec. III. Subsequently, in Sec. V, we discuss and elucidate on previous misconceptions regarding the transmission asymmetry phenomenon in the LPPPMG system class, as reported in some of the previous works discussed above. We further analyze all these aspects towards general design principles for practical set-ups.  Finally, we present our conclusions in Sec. VI and discuss the relevance of this work to current and emerging photonic applications.   
\par           
\section{The metagrating paradigm system and the onset of transmission symmetry breaking} 
\par
Diffraction gratings are widely-researched \cite{russell, zengerle} popular optical components. A diffraction grating embedded in vacuum with interfaces cut along its periodic direction with a pitch $a$ gives rise to higher-order reflected and transmitted beams (Bragg beams) for free-space wavelengths, $\lambdaup_{\textrm{free}}$, such that \cite{sfanom}:\\
\begin{equation}  
|\textrm{sin} \uptheta_{\textrm{inc}}+ m \frac{\lambdaup_{\textrm{free}}}{a}| <1,
\end{equation} 
for any integer $m \neq 0$, with $\uptheta_{\textrm{inc}}$ being the incident angle. Eq. (1) still yields the higher-order Bragg beams existing in vacuum even if the grating rests on a substrate at both the grating and substrate side; however, in practice if the substrate is thick and lossy some Bragg orders may disappear. The role of the substrate will be discussed more in Sec. V.
\par 
Here we consider a metagrating paradigm with its periodically-repeated building block being a three-log dielectric micropyramid lacking inversion symmetry along the propagation direction.  The permittivity value of the dielectric, $\varepsilon$, is taken to be equal to 10.6, that is representative of semiconductors at mid-infrared frequencies.  In the conceived metagrating design for either TM or TE incident light \cite{poldef} polarization rotation  is forbidden by virtue of the translational symmetry along one axis \cite{villenvue, sfpolaritonic, sfprapplied}. This means for either TM- or TE-polarized light, all reflected and transmitted beams from the metagrating, whether primary, $m=0$ in Eq. (1), or higher order, $m\neq0$ in Eq. (1), will also be TM- or TE-polarized respectively.  
\par             
\begin{figure}          
\includegraphics[width=8cm]{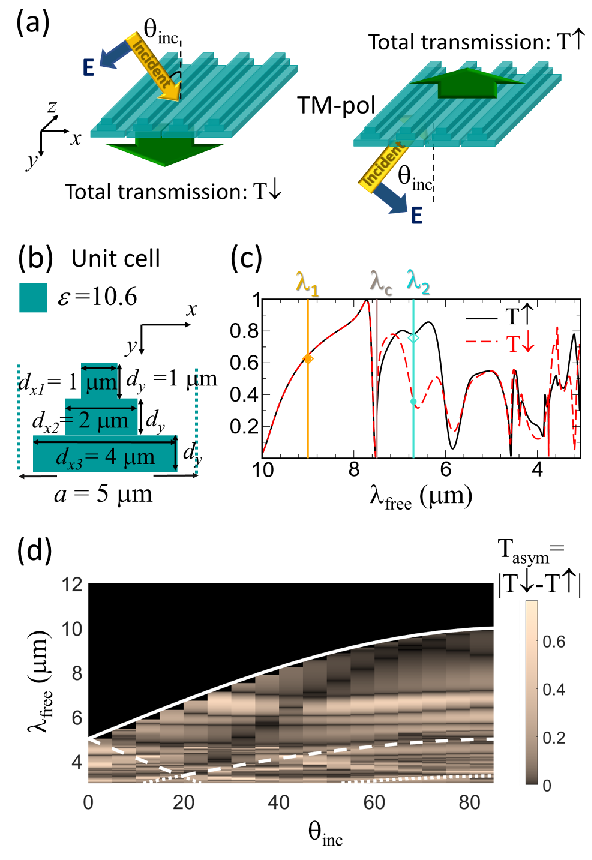}       
\caption{(a) TM-light \cite{poldef} incident from the tip side (left panel) and base side (right panel) of the micropyramid metagrating. (b) The metagrating's unit cell (translational symmetry along the $z-$ axis) (c) TMM transmission results for TM light versus free-space wavelength, $\lambdaup_{\textrm{free}}$ at $\uptheta_{\textrm{inc}}=30$ deg. Tip-side (base-side) incidence is designated as ${\textrm T}{\ignorespaces\downarrow}$ (${\textrm T}{\ignorespaces\uparrow}$) and shown with a dashed-red (solid-black) line. $\lambdaup_{\textrm 1}$, $\lambdaup_{\textrm 2}$ are indicative $\lambdaup_{\textrm{free}}$ values above and below a critical wavelength $\lambdaup_{\textrm{c}}$ where the transmission asymmetry onsets; associated results of the numerical experiment of Fig. 3 are shown with filled circles for ${\textrm T}{\ignorespaces\downarrow}$ and open diamonds for ${\textrm T}{\ignorespaces\uparrow}$.  (d) The transmission asymmetry ${\textrm T}_{\textrm{asym}}$ versus $\uptheta_{\textrm{inc}}$ and $\lambdaup_{\textrm{free}}$ (flat plotting format without interpolation); zero values indicate a symmetric response. The solid, dashed, and dotted white lines respectively represent the $\lambdaup_{\textrm{free}}$ value below which at least one, two or three higher-order Bragg diffracted beams emerge, at each side of the structure, as determined from Eq. (1).}      
\end{figure}                    
\par
\par         
We consider here the case of TM-polarized light \cite{poldef} incident at an angle $\uptheta_{\textrm{inc}}$ on such meta-grating as depicted in Fig. 1(a). Fig. 1(b) depicts the design parameters of the micropyramid's unit cell comprising the metagrating. We calculate the transmission with the transfer matrix method (TMM) \cite{TMM1, TMM2}, for $\uptheta_{\textrm{inc}}=30$ deg. and show the respective results in Fig. 1(c) versus the free-space wavelength, $\lambdaup_{\textrm{free}}$. The red-dashed line, represents the transmission, ${\textrm T}{\ignorespaces\downarrow}$, for light coming from the top side, as in the left panel of Fig. 1(a), while the black-solid line represents the transmission, ${\textrm T}{\ignorespaces\uparrow}$, for light coming from the bottom side, as in the right panel of Fig. 1(a). Interestingly, we  observe that  ${\textrm T}{\ignorespaces\downarrow}$ and ${\textrm T}{\ignorespaces\uparrow}$ exactly match for long wavelengths, and become different below a critical wavelength, $\lambdaup_{\textrm c}$. To quantify such difference we introduce the transmission asymmetry, ${\textrm T}_{\textrm{asym}}$:
\begin{equation}
{\textrm T}_{\textrm{asym}}=|{\textrm T}{\ignorespaces\downarrow} \hspace{0.1mm}-\hspace{0.1mm} {\textrm T}{\ignorespaces\uparrow}|. 
\end{equation}
We observe at $\lambdaup_{\textrm{free}}> \lambdaup_{\textrm{c}}$, ${\textrm T}_{\textrm{asym}}$ is zero designating a symmetric response, while at  $\lambdaup_{\textrm{free}} <\lambdaup_{\textrm{c}}$ we see that ${\textrm T}_{\textrm{asym}}$ varies from small to significant, taking values as large $\sim 75\%$ around $\lambdaup_{\textrm{free}} \sim$ 3.2 $\mu$m.  
\par    
We aim to understand what is the physical reason behind such a critical wavelength, $\lambdaup_{\textrm{c}}$, below which transmission symmetry breaking onsets. Clearly, polarization rotation enabling transmission asymmetry in various reciprocal systems \cite{carsten, halfswas,twistlight, kenanakis, povinelli, asymyogesh, dielrot, ewold} cannot be the responsible mechanism since it is forbidden here \cite{villenvue, sfpolaritonic, sfprapplied}. To gain more insight, we calculate the transmission asymmetry, ${\textrm T}_{\textrm{asym}}$, versus the free-space wavelength and incident angles. We start from normal incidence with a step of 5 deg. until near-grazing incidence and show the results in Fig. 1(d). We observe an abrupt separation of the ($\uptheta_{\textrm{inc}}$, $\lambdaup_{\textrm{free}}$) space into two areas with corresponding zero, and non-zero values of ${\textrm T}_{\textrm{asym}}$.   
\par      
If we look closer at these limiting $\lambdaup_{\textrm{free}}$ values for certain $\uptheta_{\textrm{inc}}$ at the boundary between these two aforementioned areas, we find that astonishingly, these are tracing the respective values where at least one higher-order Bragg diffracted beam emerges, as determined from Eq. (1). We represent the latter in Fig. 1(d) with the solid-white line. There may be more than one higher-order Bragg beam. Conversely, dashed-white and dotted-white lines represent respectively the $\lambdaup_{\textrm{free}}$ values at different $\uptheta_{\textrm{inc}}$, determined from Eq. (1), below which at least two or three higher-order Bragg beams emerge. Indeed, it is the emergence of the very first Bragg beam with $m=-1$ that abruptly changes the behavior of the system from having a symmetric to having an asymmetric transmission response. Such first-order Bragg beam is present on both sides of the structure of Fig. 1. We note in passing while Fig. 1 depicts the TM-polarization results, we have also verified that asymmetry abruptly kicks-in when the first Bragg beam with $m=-1$ emerges for the case of TE-polarized light as well \cite{poldef}. 
\par 
In the following, we consider a general LPPPMG system and demonstrate why the emergence of at least one higher-order Bragg beam enables transmission asymmetry of linearly polarized light without violating reciprocity.  
\vspace{0.05cm}    
\par              
\section{Bragg beams as channels of a reciprocal multi-port network: the route to transmission symmetry breaking} 
\par 
We present here an analogy between Bragg-diffracted beams at the surface of a LPPPMG system and networked ports. For gratings resting on a substrate the LPPPMG system would comprise both the grating and the substrate \cite{structnote}. In other words, the ports are always defined in vacuum, where transmission is measured in experiments, with each higher-order Bragg beam contributing two ports, one for the reflection and one for the transmission channel. This means a two-port system, as seen in Fig. 2(a), would represent cases where no higher-order out-coupled Bragg beams may exist, such as the metagrating system for ($\lambdaup_{\textrm{free}}$, $\uptheta_{\textrm{inc}}$) values above the white solid line of Fig. 1(d). Conversely, a four-port system [Fig. 2(b)] describes cases where one higher-order Bragg beam emerges, such as the metagrating system for ($\lambdaup_{\textrm{free}}$, $\uptheta_{\textrm{inc}}$) values between the white solid and dashed lines of Fig. 1(d) while a six-port system corresponds to cases with two higher-order Bragg beams, such as the metagrating system for ($\lambdaup_{\textrm{free}}$, $\uptheta_{\textrm{inc}}$) values between the white dashed and dotted lines of Fig. 1(d) and so on and so forth.  
\begin{figure}            
\includegraphics[width=8cm]{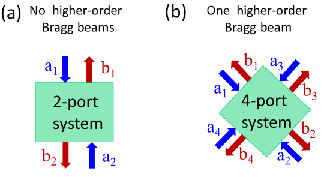}   
\caption{Multi-port network representing scattering of a linearly-polarized plane-wave from the LPPPMG metagrating system. $|\textrm{a}_i|^2$ ($|\textrm{b}_i|^2$) yields the respective beam's power through the $i^{th}$ port into (out-of) the system. Beams into and out-of all ports are depicted although some may have a zero associated coefficient. For incidence from the top [bottom] side, Port (1) [(2)] denotes the channel of the the input beam and the only or primary reflected beam, Port (2)[(1)] the channel of the only or primary transmitted beam, Port (3) [(4)] the channel of the first-order Bragg reflected beam and Port (4) [(3)] the channel of the first-order Bragg transmitted beam.}  
\end{figure}                 
\par    
We have that $[{\textrm b}]=S [{\textrm a}]$, with $[{\textrm a}]$, $[{\textrm b}]$ being column arrays contain the respective coefficients $a_i$ and $b_i$ yielding the power throughput of the incoming and outgoing beams at the $i^{th}$ port when their modulus squared is taken (see Fig. 2). $S$ is the scattering matrix, which must be symmetric for a reciprocal system \cite{caloz, tutorial, pozar},  i.e. $S=S^{\textrm T}$, with $\textrm T$ denoting the matrix transpose. Then, after adopting the notation $S_{ij}$ to represent the scattering matrix element from the $j^{th}$ to the $i^{th}$ port we obtain for the two-port system:
\begin{equation} 
{\textrm T}{\ignorespaces\downarrow}={\frac{|{\textrm b}_2|^2}{|{\textrm a}_1|^2}}\bigg\rvert_{\textrm a_2=0}=|S_{21}|^2, \hspace{1mm} \textrm{and,}\nonumber
\end{equation}
\begin{equation} 
{\textrm T}{\ignorespaces\uparrow}={\frac{|{\textrm b}_1|^2}{|{\textrm a}_2|^2}}\bigg\rvert_{\textrm a_1=0}=|S_{12}|^2. 
\end{equation}     
\par  
We see from Eq. (3) that reciprocity mandates ${\textrm T}{\ignorespaces\downarrow}$  and ${\textrm T}{\ignorespaces\uparrow}$ to be equal. Hence, transmission must be symmetric for LPPPMG systems where no higher-order Bragg beams emerge, which effectively behave as a two-port system. If additionally these systems are lossless, such as the metagrating paradigm of Fig. 1, reflection must also be symmetric. On the other hand, if these systems have a loss, gain or both, then they  may exhibit an asymmetric reflection   which however must be offset by an equivalent asymmetry in absorption or gain, thus resulting in a symmetric transmission (e.g. see Refs. \cite{sfprapplied, icton, potton, kottos, veronis2, berini}). 
\par    
Conversely, for the four-port system [case of Fig. 2(b)], taking into account that there is no input power in some ports, we obtain:
\begin{equation} 
{\textrm T}{\ignorespaces\downarrow}={\frac{|{\textrm b}_2|^2+|{\textrm b}_4|^2}{|{\textrm a}_1|^2}}\bigg\rvert_{{\textrm a_2=0},{\textrm a_3=0},{\textrm a_4=0} }=|S_{21}|^2+|S_{41}|^2, \textrm{and}, \nonumber 
\end{equation}    
\begin{equation} 
{\textrm T}{\ignorespaces\uparrow}={\frac{|{\textrm b}_3|^2+|{\textrm b}_1|^2}{|{\textrm a}_2|^2}}\bigg\rvert_{{\textrm a_1=0},{\textrm a_3=0},{\textrm a_4=0} }=|S_{32}|^2+|S_{12}|^2. 
\end{equation}    
\par    
Hence, for a LPPPMG system behaving as a four-port system, such as the metagrating system of Fig. 1 when one high-order Bragg beam emerges, transmission will be in general asymmetric with an asymmetry equal to:
\begin{equation} 
{\textrm T}_{\textrm{asym}}=\left||S_{21}|^2+|S_{41}|^2-|S_{32}|^2-|S_{12}|^2\right|=\nonumber
\end{equation}       
\begin{equation}  
\left||S_{41}|^2-|S_{32}|^2\right|,     
\end{equation}  
This is because while reciprocity requires $S=S^{\textrm T}$ and thus $S_{21}$ to be equal to $S_{12}$, it does not require $S_{41}$ and $S_{32}$ to be equal. In a similar manner we find that transmission is generally asymmetric for a LPPPMG system with two higher-order Bragg beams, behaving as a six-port system, and so on and so forth. 
\par    
We note in passing that the higher-order Bragg channels on one of the structure's sides may be engineered to have no output, e.g. like in the case of systems studied in Ref. \onlinecite{lockyear}. In the latter cases, the higher-order channels on one of the structure's side are inactive/closed, i.e. effectively non-existent with a zero corresponding scattering matrix element $S_{ij}$. Engineering this behavior is rather challenging, with designs being generally less practical, especially in the infrared and optical spectrum, without necessarily a higher asymmetry performance. We discuss these type of cases in more detail in Sec. V-C. The important take away of this section is, as Eq. (5) attests, that it just takes the emergence of only one higher-order Bragg beam to break transmission symmetry, whether the corresponding channels of such higher-order beam are open on either or both sides of the structure. However, in general, wherever any higher-order Bragg beams emerge these will act as open channels on both sides of the structure, like in the case of our metagrating paradigm of Fig. 1.
\par
In the following,  we perform a first-principle numerical experiment with the metagrating of Fig. 1 at $\lambdaup_{\textrm{free}}=\lambdaup_2$, demonstrating  how transmission symmetry breaking occurs by means of the first-order transmitted Bragg beams ($m=-1$) that can out-couple from both structure sides, albeit with different strengths, $|S_{41}|^2$ and $|S_{32}|^2$, respectively.
\par    
\begin{figure}[!htb]               
\includegraphics[width=8cm]{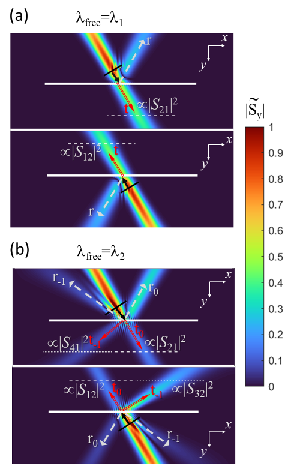}  
\caption{Numerical experiment (FDTD) of TM light impinging at 30 deg. at the metagrating (white rectangle) for  the wavelengths $\lambdaup_1$ [case shown in (a)] and $\lambdaup_2$ [case shown in (b)] designated in Fig. 1(c). The $y$-component of the normalized time-averaged Poynting vector (absolute value) is depicted ($\vert\tilde{\textrm{S}}_y\rvert$). The top (bottom) panels in (a) and (b) correspond to incidence from the tip (base) side of the micropyramid array. The slanted-black line designates the source launching the incident Gaussian beam, marked with the black solid arrow. The white dashed and red double arrows represent respectively the reflected (marked with r) and transmitted (marked with t) beams, with the subscripts in (b) designating diffraction order. The dashed and dotted horizontal line segments in all panels represent the respective ports; the corresponding total power throughput would be proportional to $|S_{ij}|^2$, with $S_{ij}$ being the relevant Scattering matrix element. 
}        
\end{figure}                          
\par
\par        
\section{Demonstrating the mechanism of transmission symmetry breaking with a first-principle numerical experiment} 
\par
We simulate the metagrating's response to a TM quasi-monochromatic Gaussian beam \cite{quasi} launched at 30 deg. with the two-dimensional (2D) Finite-Difference Time Domain Method (FDTD) \cite{taflove} with Perfect Matched Layer (PML) boundary \cite{PML1,PML2} conditions (translational symmetry is taken out-of-the plane of incidence). We  perform the FDTD experiment for the two indicative wavelengths $\lambdaup_{\textrm{1}}$ and  $\lambdaup_{\textrm{2}}$, designated in Fig. 1(c), and show the respective results in Figs. 3(a) and 3(b) for incidence from both sides of the micropyramid-array. The micropyramid-array is represented as a white rectangle that has its tip side coinciding with the top rectangle side and its base side coinciding with the bottom rectangle side. The quantity depicted in Fig. 3 is $\vert\tilde{\textrm{S}}_y\rvert$, with $\tilde{\textrm{S}}_y$ being the normalized time-averaged $y$-component of the Poynting vector, calculated in FDTD at steady-state by averaging the $y$-component of the Poynting vector over a wave period, $T_{\textrm{per}}$, and then dividing this by the corresponding value of the source maximum.  
\par   
For $\lambdaup_{\textrm{free}}=\lambdaup_{\textrm{1}}$ we observe in Fig. 3(a) only one reflected and transmitted beam; their respective strengths appear the same irrespective of the side the incident beam is launched from. To check this further, we take the $\vert\tilde{\textrm{S}}_y\rvert$ power throughput collected over the line-segment port (dashed line) and normalize this by dividing it with the respective power throughput collected from an equivalent line-segment port at the same $y$ distance from the source, in the absence of the metagrating structure. This process yields respectively for the top and bottom panels of Fig. 3(a) $|S_{21}|^2$, and $|S_{12}|^2$; reciprocity mandates these to be equal; to no surprise $|S_{21}|^2$ and $|S_{12}|^2$ are found to be equal in the FDTD experiment \cite{numerror}. 
\par 
On the other hand, for $\lambdaup_{\textrm{free}}=\lambdaup_{\textrm{2}}$ we observe two reflected and transmitted beams; this is expected from Eq. (1) which predicts that at this wavelength and launch angle a Bragg beam of order $m=-1$ would emerge. It is interesting that we see in Fig. 3(b) that while the strength of the primary transmitted beams ($m=0$) appear the same, whether the beam is launched from the top or bottom side, the ones corresponding to Bragg order $m=-1$ have visibly very different strength. By performing, the detailed calculation of the power throughput through the dashed and dotted line ports, as we did for the case of Fig. 3(a), we confirm the initial visible observation. Indeed, the power throughput of the primary transmitted beams, corresponding to $|S_{21}|^2$ and $|S_{12}|^2$ are equal \cite{numerror}.      
\par      
However, the power though-put of the transmitted Bragg beams with order $m=-1$, corresponds to $|S_{41}|^2$ and $|S_{32}|^2$, for incidence from the top and bottom side respectively. Reciprocity does not require these to be equal and our numerical experiment reveals that in fact they do turn out to be significantly different. It is that significant difference that caused the significant transmission asymmetry we observed in Fig. 1(c). Now that we have calculated with the FDTD method the scattering matrix elements $|S_{21}|^2$ and $|S_{12}|^2$ for case of Fig. 3(a) and $|S_{21}|^2$, $|S_{12}|^2$, $|S_{41}|^2$ and $|S_{32}|^2$ for the case of Fig. 3(b), we can use them to calculate ${\textrm T}{\ignorespaces\downarrow}$ and ${\textrm T}{\ignorespaces\uparrow}$ from Eqs. (3) and (4). We do this and show the respective results with filled circles and open diamonds in Fig. 1(c); a very excellent agreement is observed with the TMM results. 
\par  
To recap, our FDTD experiment validated our hypothesis that the transmission asymmetry of linearly polarized light in the metagrating system stems from the introduction of additional transmission channels via higher-order Bragg beams.   
\par         
\section{Elucidating on previous misconceptions and practical considerations}  
\par    
We briefly discussed in the introduction previous works that have reported a transmission asymmetry of linearly polarized light incident on a LPPPMG structure. Indeed, there was apparent lack of consensus in these previous reports regarding the key responsible mechanism for such a transmission asymmetry in this class of systems. In the previous sections, we have unveiled with our systematic all-angle calculations and the multi-port network framework that the only necessary condition is the emergence of at least one higher-order Bragg beam in a LPPPMG system that lacks inversion symmetry along the propagation direction. In this section, we discuss in more detail how such previous misconceptions arose while correlating the results of these previous works with our findings here. We analyze different aspects of previous reported structures, while highlighting interpretation mistakes. The discussion below emphasizes how identifying correctly the core mechanism of transmission asymmetry of linearly polarized light in LPPPMG systems as stated above is crucial towards transferable design principles for practical implementations across the EM spectrum. 
\par        
\begin{figure}[!htb]               
\includegraphics[width=8.0cm]{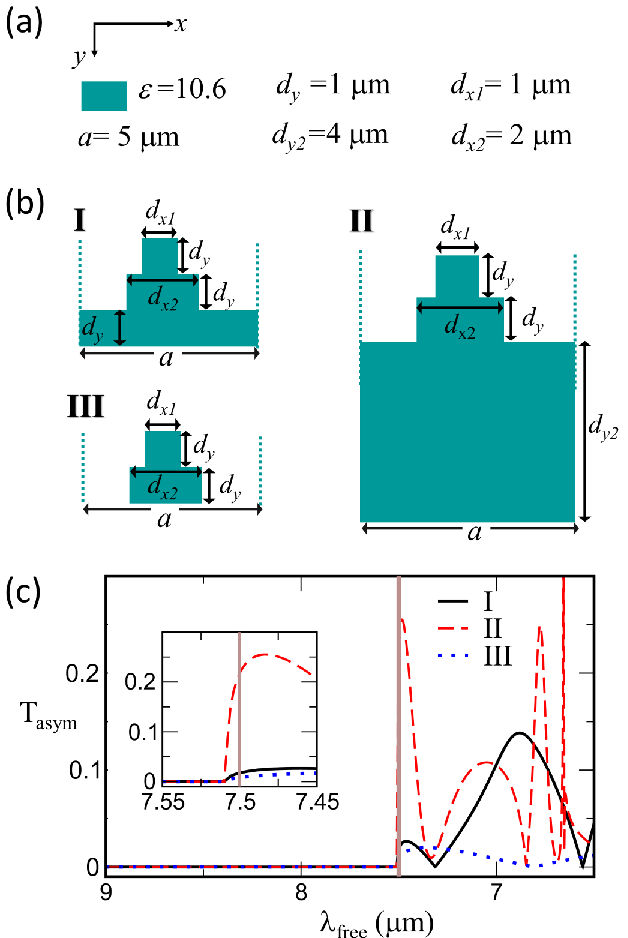}  
\caption{(a) Structural parameters of the designs in (b). (b) Unit cells for variations I, II, III of the metagrating of Fig. 1 (c) Corresponding TMM results of the transmission asymmetry, ${\textrm T}_{\textrm{asym}}$, [defined in Eq. (2)] versus free-space wavelength $\lambdaup_{\textrm{free}}$ [TM-polarized light incident at $\uptheta_{\textrm{inc}}=30$ deg]. The vertical line represents the wavelength where at least one higher-order Bragg beam emerges as determined from Eq. (1). The inset zooms around the regime of the onset of the transmission asymmetry. 
}       
\end{figure}                  
\par  
\subsection{On the substrate's role on the spectral onset of the transmission asymmetry}
\par
The design variations I, II and III of Fig. 4 aim to address the assertion of Ref. \onlinecite{aydin} that the transmission asymmetry relates to the Wood anomaly \cite{wood} and thus its spectral onset depends on the permittivity of the substrate material. If the last metagrating layer is widened until the unit cell limit, thus essentially becoming a substrate layer we obtain design I; design II has a thicker substrate layer, while in design III the substrate layer is entirely dropped [see Figs. 4(a) and 4(b) for full design parameters]. We show in Fig. 4(c) the corresponding TMM results for transmission asymmetry, ${\textrm T}_{\textrm{asym}}$, as defined from Eq. (2), for TM- light incident at $\uptheta_{\textrm{inc}}=30$ deg.. We observe the onset wavelength of the transmission asymmetry is common for all three designs, which matches the prediction from Eq. (1) seen with the brown vertical line. The tiny relative difference of $\sim 0.1\%$ is due to numerical-grid errors in TMM. 
\par        
\begin{figure}[!htb]                 
\includegraphics[width=8.0cm]{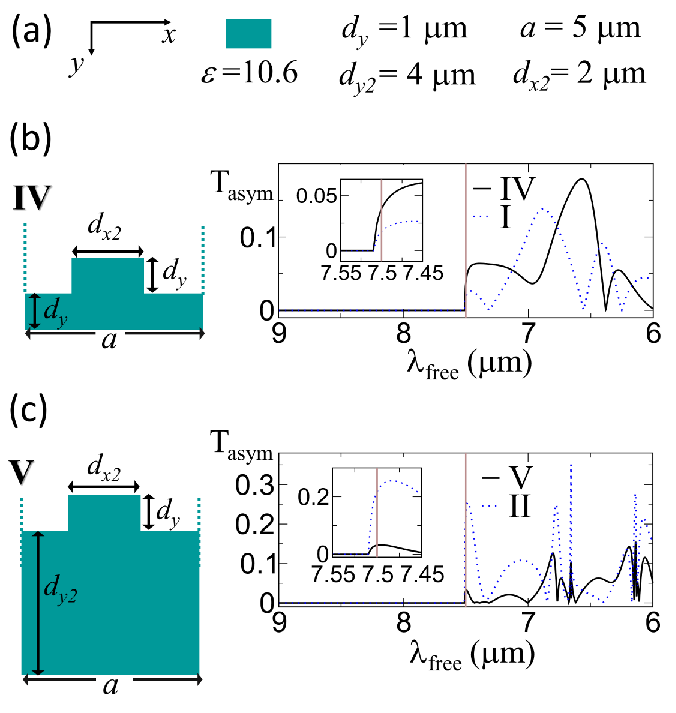}  
\caption{(a) Structural parameters of the designs in (b) and (c). (b) In the left panel the unit cell of design IV is depicted; the right panel shows the corresponding TMM result for the transmission asymmetry ${\textrm T}_{\textrm{asym}}$, [defined in Eq. (2)] versus free-space wavelength [TM-polarized light incident at $\uptheta_{\textrm{inc}}=30$ deg]. The blue-dotted line shows the corresponding ${\textrm T}_{\textrm{asym}}$ of design I for comparison. The vertical line represents the wavelength where at least one higher-order Bragg beam emerges as determined from Eq. (1). The inset zooms around the regime of transmission asymmetry onset. (c) Same as in (b) for design V, that is compared with design II.
}     
\end{figure}                   
\par 
For LPPPMG structures \cite{structnote}, the result of Fig. 4 establishes that the spectral onset of transmission asymmetry coincides with the emergence of the first higher-order beam in vacuum and is independent of the existence of any substrate or its permittivity. The erroneous conclusion of the simulation work of Ref. \onlinecite{aydin}, stating the dependence of the transmission asymmetry onset on the permittivity of the substrate, arose from transmission being calculated within the substrate layer which was taken to be  semi-infinite along the $y$-direction. Transmission is defined however and indeed measured in experiments through the substrate in vacuum; thus transmission should be calculated in vacuum after the substrate as we have done for the designs I and II that comprise a substrate layer. Note, that the structures of Ref. \onlinecite{aydin} as well as all structures studied here have been taken to be infinitely extending in the $x$-direction. We alert the reader that the unavoidable finite lateral extent of the structure may also come into play in experiments, that may affect the transmission asymmetry. The cause of such lateral finite-size effects is the possible beam leakage from the side structural faces (along the $yz$ plane), which would be more likely to occur for thick substrates. So, when designing practical systems such lateral finite-size effects need to be taken into account.  
\par           
\subsection{On the substrate's role on structural inversion symmetry breaking along the propagation direction }
\par    
We established with Fig. 4 that adding a substrate on a LPPPMG structure comprising structural elements lacking inversion symmetry along the propagation direction ($y$ direction here) has no effect on the transmission asymmetry onset. Now, structures that possess inversion symmetry along the propagation direction, would always show a symmetric transmission response. For example, transmission would be always symmetric for a structure resulting from design III of Fig. 4 after e.g. dropping the top layer, as the resulting structure would have $y$-inversion symmetry. Things would change however if we were to place the aforementioned structure on a substrate. The substrate itself breaks the $y$-inversion symmetry, without requiring the structural elements resting on the substrate to be lacking $y-$inversion symmetry themselves. 
\par    
To demonstrate this, we consider here designs IV and V, which are systems comprising the same $y-$symmetric  structural element, a rectangular-cross-section log with sides of $d_{x2}=$2 $\mu$m, along the $x-$ direction and  $d_y=$1 $\mu$m, along the $y-$ direction and infinite in the $z-$ direction. Such a log is resting on a substrate made of the same dielectric material, with permittivity $\varepsilon=$10.6, and being 1 $\mu$m thick for design IV and 4 $\mu$m thick for design V. The corresponding results for the transmission asymmetry, ${\textrm T}_{\textrm{asym}}$, defined in Eq. (2),  are shown in Fig. 5, for the case of TM-light incident at a 30 deg. angle. Indeed, these results attest for a broken transmission symmetry for wavelengths below the value of 7.5 $\mu$m that is designated with the brown vertical line in Fig. 5 and coincides with the wavelength where a higher-order Bragg beam with $m=-1$ emerges. 
\par    
Note that Ref. \onlinecite{optcom} which we discussed in the introduction, reported transmission asymmetry for a structure comprising silver rectangular logs that are symmetric along the propagation direction and rest on a dielectric substrate. However, Fig. 5 additionally reveals that the substrate can introduce significant asymmetry in the transmission response even in a fully dielectric system and even if the substrate material and the material comprising the periodic structural elements are the same. In the context of designs for practical systems, a key question is how important is it to have a $y-$asymmetry for the structural element alone that rests on the substrate.    
\par   
To answer this we compare design IV with design I and design V with design II. The respective design pairs are the same in every respect except for an extra top square log, with a  side of 1 $\mu$m, that is present in designs I and II and makes them $y-$asymmetric even without the substrate. The corresponding ${\textrm T}_{\textrm{asym}}$ of designs I and II is shown in Fig. 5 with a blue-dotted line. The results in Fig. 5 reveal that there is no rule of thumb. In fact, Fig. 5 shows that $y-$asymmetric structural elements on a substrate may or may not yield a better transmission asymmetry performance in comparison to $y-$symmetric ones; it depends on the particulars of the structure and the frequency. However, $y-$asymmetric structural elements on a substrate most certainly offer a broader parameter space for design flexibility towards targeted applications in comparison to $y-$symmetric ones.  
\par              
\subsection{Different number of higher-order Bragg beams at each side of the structure and transmission symmetry breaking}
\par   
The all-dielectric meta-grating systems of Figs. 1 and 3-5 have the same number of Bragg beams at both sides of the structure \cite{structnote}. This would in general be true for most all-dielectric systems. Actually, it takes quite complex engineering to suppress higher-order Bragg beams on one side of LPPPMG systems only and would certainly typically involve material loss. So for normal incidence, the typical situation would involve two higher-order beams with $m=-1$ and $m=1$ at each interface side for frequencies above the first Bragg condition but below the second. Hence, as we discussed in Sec. III, the typical situation would involve a six-port system [see Fig. 6(a)]. However, Ref. \onlinecite{lockyear} considered a meta-grating structure comprising connected air-waveguides in a perfect metal background leading to interfaces having a different pitch. The different pitch at each interface along with the forbidden propagation inside the metal made it possible to inactivate the higher-order Bragg beam channels on one of the metagrating's sides only in this particular system. Thus, this system may be described by the four-port system seen in Fig. 6(b) [instead of that of Fig. 6(a)].    
\par        
The authors of Ref. \onlinecite{lockyear} concluded that the transmission asymmetry occurs because of the emergence of higher-order beams on one side of the structure only. Indeed, this is a special case that yields transmission asymmetry with:\\ 
\begin{equation}
{\textrm T}_{\textrm{asym}}=\left||S_{41}|^2+|S_{61}|^2\right|.   
\end{equation} 
However, the structure would have still yielded asymmetry in transmission even if higher-order beams where present on both sides of the interface [situation in Fig. 6(a)] given by: 
\begin{equation} 
{\textrm T}_{\textrm{asym}}=\left||S_{41}|^2+|S_{61}|^2-|S_{32}|^2-|S_{52}|^2\right|. 
\end{equation} 
\par  
Not only it is not a requirement to have a different number of Bragg channel at each side of the structures for transmission asymmetry, but also there is no guarantee this route would give rise to a higher asymmetry either. One should not be misled by Eq. (6) and (7) to conclude that transmission asymmetry is necessarily larger when there is a different number of available Bragg channels at each side of the structure; the latter corresponding systems typically would include material loss or high reflection which limits the attainable $|S_{41}|$, $|S_{61}|$ values. Indeed, the reported transmission asymmetry in the metallic structure of Ref. \onlinecite{lockyear}, where two higher-order Bragg beams exist on one side only, is about 30$\%$. As an example, we highlight the case of the all-dielectric metagrating paradigm of Fig. 1 for TM-polarization and 30 deg. incident angle, where we were able to obtain a transmission asymmetry as high as $\sim$75$\%$ at about 3.2 $\mu$m. Two higher-order Bragg channels (with $m=-1$ and $m=-2$) are present at both sides of the structure in this example.
\begin{figure}               
\includegraphics[width=8cm]{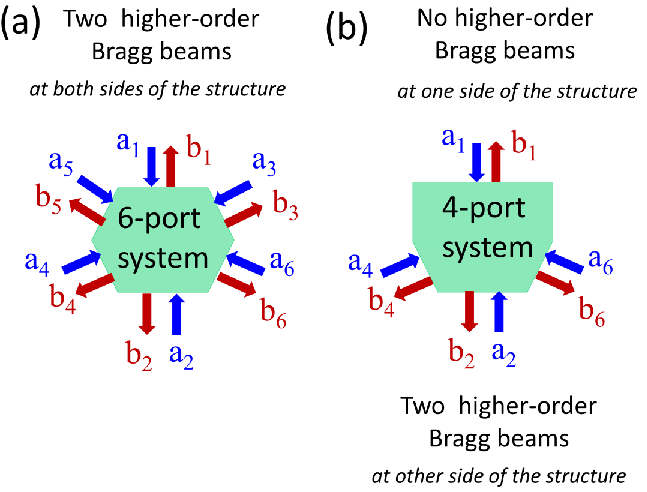}  
\caption{Multi-port network representions, like in Fig. 3, but for cases: (a) Where two higher-order Bragg beams exist on both sides of the structure and (b) Two higher-order Bragg beams exist on one side of the structure but no higher-order Bragg beams exist on the other side. The $\textrm{a}_i$, $\textrm{b}_i$ coefficients have the same meaning as in Fig. 2. Here, for incidence from the top [bottom] side, Port (1) [(2)] denotes the channel of the input beam and the only or primary reflected beam, Port (2)[(1)] the channel of the only or primary transmitted beam, Ports (3), (5) [(4), (6)] the channels of the Bragg reflected beams and Ports (4), (6) [(3), (5)] the channels of the Bragg transmitted beams. Ports (3) and (5) do not exist for case (b), where two higher-order beams are present at one side of the structure only, like in the system reported in Ref. \onlinecite{lockyear}.}    
\end{figure}                        
\par   
To resume, while having a different number of Bragg beams at each of the structure sides may lead to transmission asymmetry this is not a necessary condition and it does not guarantee a stronger transmission-asymmetry performance either. Additionally, designs that can give a different number of diffraction orders at each side are considerably more complex, and likely needing material loss; these complex designs are not as easily realizable for infrared and optical frequencies. Hence, cost-benefit considerations certainly do not justify pursuing these complex designs for transmission asymmetry in practical applications at these frequencies. In other words, the design focus to obtain transmission asymmetry in LPPPMG systems of interest in infrared and optical applications should be towards systems with at least one higher-order Bragg channel existing at both sides of the structure \cite{structnote}.    
\par  
\subsection{On the role of the pitch and incorporating structural elements of different pitch for transmission symmetry breaking}  
\par    
In this sub-section we discuss in more detail the role of the pitch, i.e. the lateral periodicity in LPPPMG systems, including in cases of LPPPMG systems where structural elements of a different lateral pitch have been incorporated. 
\par     
Tunability, i.e. spectral shift, of the transmission asymmetry onset by changing the pitch of the structure has been previously observed \cite{hamza2}, albeit with no explanation why this occurs, --although also in this work transmission is calculated within a semi-infinite substrate, which means the spectral location of the transmission asymmetry onset has been incorrectly identified for the same reason as in Ref. \onlinecite{aydin}. However, indeed, it is the pitch $a$ that control the spectral onset of the transmission asymmetry because it {\it does} control the spectral emergence of the first Bragg beam; per Eq. (1) the important parameter is $\lambdaup_{\textrm{free}}/a$. So, by changing $a$, the first higher-order Bragg beam at a certain incident angle, would emerge at a different free-space wavelength, $\lambdaup_{\textrm{free}}$. For example, in all metagrating systems of Figs. 1, 3, 4 and 5, for an incident angle of 30 deg. transmission asymmetry onsets at $\lambdaup_{\textrm{free}}= 1.5 a$. 
\par   
However, Ref. \onlinecite{concita} additionally asserted that transmission symmetry breaking occurs because a wavelength smaller than the pitch allows the impinging wave to ``see'' the 2D structuring of the system. Indeed $a/\lambdaup_{\textrm{free}}=1$ coincides with the Bragg condition at normal incidence. However, the cause for the transmission asymmetry is the mere emergence of the higher-order Bragg beams. It is the emergence of the additional Bragg beam(s) that allows preserving reciprocity while breaking transmission symmetry as we discussed in Sec. III. The assertion of Ref. \onlinecite{concita} that a wavelength smaller than the pitch is needed for the wave to ``see'' the structure for transmission asymmetry to emerge is certainly not correct and not consistent with an all-angle analysis as we have done here. Actually, as we observe in Fig. 1 for near-grazing incident angles transmission symmetry breaks at free-space wavelength values almost twice the pitch. 
\par 
Also, there was a previous report of transmission asymmetry in an LPPPMG system that included structural elements of different lateral periodicity \cite{cascgrat}. For such systems it is the larger pitch that controls the spectral onset of transmission asymmetry. For example, at normal incidence transmission asymmetry onsets at $\lambdaup_{\textrm{free}}=a$, with $a$ here being the periodicity of the larger pitch, which is consistent with the results of Ref. \onlinecite{cascgrat}. In the systems of Ref. \onlinecite{cascgrat} the role of structural elements of different pitch basically just boils down to breaking inversion symmetry along the propagation direction. However, if such inversion symmetry was broken by any other way without having elements of different periodicity, the system would still show an asymmetric transmission whenever higher-order Bragg beams exist beyond the structure's interfaces \cite{structnote}. In other words, while incorporating elements of different periodicity is a route to breaking inversion along the propagation direction, this is not the only route and certainly not the most practical. Accordingly, there is no apparent advantage in targeting LPPPMG designs comprising elements with a different lateral periodicity.
\par  
\section{Conclusions}    
\par  
With all-angle transmission calculations and the aid of a multi-port network analysis, we showed that a reciprocal route to transmission asymmetry of linearly-polarized light without rotation of its polarization is to excite at least one higher-order Bragg diffracted beam on either or both sides of a LPPPMG structure \cite{structnote} that has broken inversion symmetry along the propagation direction. We have further demonstrated this core mechanism in a metagrating paradigm with a numerical experiment. Moreover, we established that asymmetric transmission can occur even when the same number of diffracted beams are present at both sides of the structure, contrary to previous reports asserting the need for a different number \cite{lockyear}. We further disproved the conclusion of Ref. \onlinecite{aydin} and showed that the onset wavelength of the transmission asymmetry is actually independent on the presence of a substrate. Finally, our results and analysis revealed that complex designs incorporating structural elements of different pitch are neither necessary nor have an apparent advantage.
\par           
Thus, this work offers transferable design insights for transmission symmetry breaking of linearly polarized light across the EM spectrum with simple LPPPMG systems that lack inversion symmetry along the propagation direction. The mechanism uncovered here for transmission asymmetry would be applicable to linear passive periodic metasurfaces \cite{yuribronger} as well, however the contribution from polarization rotation that is possible in these systems would need to be accounted as well. We believe this work will further inspire non-linear self-biased non-reciprocal \cite{caloz, engheta} metasurfaces and metagratings that are highly relevant to applications like integrated infrared photonics \cite{juejun}, self-isolating lasing architectures \cite{dionne} and directional second-harmonic generation \cite{yuri2,lniob}. Other interesting asymmetric effects that have been reported involve the lack of transmission mirror symmetry with respect to the surface normal caused by the excitation of topological surface waves \cite{aklesh}, i.e. a transmission that is different for incident angles $\uptheta$ and $-\uptheta$. Hence, the metagrating systems reported here may further inspire functional designs for highly versatile light flow control with both mirror and propagation-direction transmission asymmetry.     
\section*{Acknowledgements}      
We would like to thank the UNM Center for Advanced Research Computing, supported in part by the National Science Foundation, for providing the high performance computing resources used in this work. 
\par   
 

\begin{thebibliography}{99}
\bibitem{passive}
In the context of this paper, the term ``passive'' means systems without an external magnetic field, without time-modulation of the optical responses of its constituents and without gain.
\bibitem{note1}  
Planar asymmetry is sufficient without requiring chirality \cite{nanoph_tretyak}.
\bibitem{poldef}  In TM-(TE-) polarization, the electric (magnetic) field lies on the plane of incidence which is normal to the  dimension of translational symmetry; the magnetic (electric) field is parallel to the  dimension of translational symmetry (standard Optics definition); this is opposite to early polarization labels common in the field of two-dimensional photonic crystals \cite{joannbook}.
\bibitem{structnote} Note, the term ``structure'' is used here to include any substrate. The Bragg beams in the discussions in this paper are always the beams present in vacuum, beyond the substrate layer when present.
\bibitem{numerror} We find a small relative difference of less than $0.01\%$ due to numerical grid errors.
\end{thebibliography}
\end{document}